# Ti-alloying of BaZrS$_3$ chalcogenide perovskite for photovoltaics

Xiucheng Wei[a], Haolei Hui[a], Samanthe Perera[a], Aaron Sheng[b], David F. Watson[b], Yi-Yang Sun[c], Quanxi Jia[d], Shengbai Zhang[e], and Hao Zeng[a,*]

[a]Department of Physics, University at Buffalo, the State University of New York, Buffalo, NY 14260, USA

[b]Department of Chemistry, University at Buffalo, the State University of New York, Buffalo, NY 14260, USA

[c]State Key Laboratory of High Performance Ceramics and Superfine Microstructure, Shanghai Institute of Ceramics, Chinese Academy of Sciences, Shanghai 201899, China

[d]Department of Materials Design and Innovation, University at Buffalo, the State University of New York, Buffalo, NY 14260, USA

[e]Department of Physics, Applied Physics & Astronomy, Rensselaer Polytechnic Institute, Troy, NY 12180, USA

* corresponding author: haozeng@buffalo.edu

**ABSTRACT**

BaZrS$_3$, a prototypical chalcogenide perovskite, has been shown to possess a direct band gap, an exceptionally strong near band edge light absorption, and good carrier transport. Coupled with its great stability, non-toxicity with earth abundant elements, it is thus a promising candidate for thin film solar cells. However, its reported band gap in the range of 1.7-1.8 eV is larger than the optimal value required to reach the Shockley-Queisser limit of a single junction solar cell. Here we report the synthesis of Ba(Zr$_{1-x}$Ti$_x$)S$_3$ perovskite compounds with a reduced band gap. It is found that Ti–alloying is extremely effective in band gap reduction of BaZrS$_3$: a mere 4 at% alloying decreases the band gap from 1.78 to 1.51 eV, resulting in a theoretical maximum power conversion efficiency of 32%. Higher Ti-alloying concentration is found to destabilize the distorted chalcogenide perovskite phase.

**KEYWORDS**: chalcogenide perovskite, band gap, alloying, Shockley-Queisser limit, stability

**INTRODUCTION**

Research in organic-inorganic hybrid perovskites has experienced phenomenal progress.[1-7] The power conversion efficiency (PCE) of halide perovskite solar cells has witnessed an extremely rapid increase, from an initial PCE of 3.8% in 2009[4] to above 25% in 2019.[8] The extremely long carrier diffusion length originates from its defect tolerance.[9-10] However, halide perovskites suffer from intrinsic instability from exposure to air, heat and electric field.[11] As the material is formed at 100 °C, its low formation energy barrier is also what contributes to its intrinsic instability.[12-14] Its sensitivity to moisture stems from the strong ionicity.

To address the stability and toxicity issues of halide perovskites, we and others proposed and synthesized chalcogenide perovskites as an alternative family of materials, with the intention to replace halide perovskites for photovoltaic and other optoelectronic applications.[15-19] Intense green luminescence was found in both undoped and heavily n- and p-doped $SrHfS_3$, suggesting important applications for efficient green Light emitting diodes.[20] On the theory front, Ruddlesden–Popper perovskite sulfides $A_3B_2S_7$ were proposed as a ferroelectric photovoltaic materials in the visible regime.[21] Agiorgousis *et al.* isolated $Ba_2AlNbS_6$, $Ba_2GaNbS_6$, $Ca_2GaNbS_6$, $Sr_2InNbS_6$, and $Ba_2SnHfS_6$, screened from hundreds of double perovskites, as the most promising photovoltaic materials.[22] Jaramillo *et al.* provided a strategy for discovering highly polarizable semiconductors including chalcogenide perovskites.[23] Another recent theoretical paper predicted that 2D $Ca_3Sn_2S_7$ possess a graphene-like linear electronic dispersion with a direct band gap and ultrahigh carrier mobility comparable to that of graphene.[24]

Among the chalcogenide perovskites, the most studied material is $BaZrS_3$. Most recently, we have shown that $BaZrS_3$ thin films possess exceptionally strong light absorption with near edge absorption coefficient larger than $10^5$ cm$^{-1}$.[18] Such material also shows reasonable room temperature mobility with a value ranging from 2.3 to 13.7 cm$^2$/Vs depending on processing conditions.[18] However, its band gap is less optimal for a single junction solar cell. With a band gap of 1.74 eV, the theoretical PCE using BaZrS$_3$ as a light absorber is 28%. It has been shown theoretically that its band gap can be reduced by alloying with 3d cation Ti to lower the conduction band minimum (CBM) or with 4p anion Se to uplift the valence band maximum (VBM).[16] In this work we prepared powder samples of Ba(Zr$_{1-x}$Ti$_x$)S$_3$ chalcogenide perovskite compounds, with x ranging from 0 to 0.1. We show that small concentration of Ti alloying results in very effective band gap narrowing: a mere 4 at% alloying decreases the band gap of BaZrS$_3$ from 1.78 to 1.51 eV, resulting in a theoretical maximum PCE of 32% for single junction solar cells. Higher Ti-alloying concentration, however, is found to destabilize the distorted chalcogenide perovskite phase. Our results suggest that Ba(Zr$_{1-x}$Ti$_x$)S$_3$ materials are promising candidates as solar absorbers. A similar work was published recently,[25] showing a band gap reduction of 300 meV for Ba(Zr$_{0.95}$Ti$_{0.05}$)S$_3$, although the reported band gap of both BaZrS$_3$ and Ti-alloyed one is about 150 meV higher than ours and other published work.[15-17]

## RESULTS AND DISCUSSIONS

Typical SEM images of Ba(Zr$_{1-x}$Ti$_x$)S$_3$ powder samples with x = 0 and 0.04 are shown in Figures 1(a) and 1(b), respectively. Both images suggest that the Ba(Zr$_{1-x}$Ti$_x$)S$_3$ powder has a grain size of a few microns, indicating no crystallinity degradation upon Ti alloying. In Figure 1(c), the EDX analysis for pure BaZrS$_3$ powder shows a composition of Ba: Zr: S ratio of 21.7: 21.2: 57.1. The slight deviation from perfect 1: 1: 3 ratio may result from slight sulfur vacancies due to the high processing temperatures.[15] In Figure 1(d), the EDX analysis for 4 at% Ti alloyed Ba(Zr$_{1-x}$Ti$_x$)S$_3$ powder shows a composition ratio of Ba: Zr: Ti: S = 21.8: 21.3: 0.8: 56.1. The measured Ti composition may not be accurate due to very low concentration of Ti as well as the overlap of energies between Ba Lα and Ti Kα characteristic x-ray emission lines. Nevertheless, the presence of Ti can be clearly observed from the different peak shape at around 4.5 keV, shown in the insets of Figures 1(c) and (d).

X-ray diffraction (XRD) results of Ti alloyed Ba(Zr$_{1-x}$Ti$_x$)S$_3$ powder samples for x=0, 0.01, 0.02, 0.03, and 0.04 are shown in Figure 2(a). All peaks match with the standard PDF card JCPDS 00-015-0327, suggesting an orthorhombic distorted perovskite structure with *Pnma* space group. The enlarged view of the first three strongest peaks (121), (040) and (240) of the Ba(Zr$_{1-x}$Ti$_x$)S$_3$ can be seen in Figures 2(b)-(d), respectively. A systematic peak shift is observed for all the three peaks. The peak positions for unalloyed and 4 at% Ti alloyed samples are indicated by vertical dotted lines, with a difference of 0.07° between each pair of dotted lines. Such shift can also be observed in Ba(Zr$_{1-x}$Ti$_x$)O$_3$ powders as shown in Figure S1. It can be seen that as titanium alloying level increases, the peaks shift to larger diffraction angles, indicating a decrease in d spacing. The value of d$_{121}$ as a function of Ti alloying percentage is plotted in Figure 2(e), derived from Bragg's law. The systematic decrease of lattice spacing with increasing Ti concentration suggests that the Ti element is incorporated in the perovskite lattice, instead of forming interstitials or secondary phases. At higher Ti alloying percentage of x = 0.05, 0.075 and 0.1 (Figure S3), secondary phases start to show up. These secondary phases are identified as binary BaS, TiS$_2$, ZrS$_2$ and ternary BaTiS$_3$. These results suggest that at Ti concentration higher than 5 at%, the Ti-alloyed BaZrS$_3$ perovskite phase is not stable, and phase separation occurs. This is consistent with earlier theoretical predictions.[16, 26]

The Raman spectra for Ba(Zr$_{1-x}$Ti$_x$)S$_3$ with different Ti concentration measured at room temperature are shown in Figure 3. Several Raman modes can be identified. No obvious peak shift at different Ti concentrations can be seen for the Raman modes $B_{1g}^1$, $A_g^4$, $B_{2g}^6$, and $B_{1g}^4$, possibly due to the instrumental limit. We observed no additional Raman modes for Ti alloying concentration less than or equal to 4 at%, indicating no phase separation at this level. It, however, is interesting to observe a systematic peak shift from 391 to 404.4 cm$^{-1}$, accompanied by a strong increase in its intensity, with increasing x from 0 to 0.04. Our previous work suggests that the forbidden LO-phonon scattering is the most likely cause of the 390-440 cm$^{-1}$ anomaly for BaZrS$_3$ observed using 676 nm resonant excitation, corresponding mainly to Zr-S stretching modes[23]. In this work, the 514 nm off-resonance excitation leads to very weak features in this wavenumber region for BaZrS$_3$. However, with Ti substitution of the Zr-sites, the selection rules can be relaxed due to the structure disorder, such as modified bond length and angle. Therefore, an increase in intensity together with a shift of the peak position to higher wavenumbers with increasing Ti-alloy concentration clearly indicates that Ti is substitutionally alloyed into the Zr-site of the perovskite lattice. However, as Ti alloying concentration further increases to x = 0.075 and 0.1, more peaks in the 350-500 cm$^{-1}$ range start to emerge (Figure S4), which may come from phonon scattering of the secondary phases. This is consistent with the XRD results, showing additional peaks from secondary phases at higher Ti concentration.

The band gap of BaZrS$_3$ has been reported to be between 1.7 to 1.8 eV.[15-16, 27-29] However, it is higher than the optimal value required to reach the Shockley-Queisser limit of a single junction solar cell.[30] We characterized the band gap of Ba(Zr$_{1-x}$Ti$_x$)S$_3$ using UV-vis absorption spectroscopy, and extracted the band gap using Tauc plot. Figure 4(a) shows the Tauc plot of Ba(Zr$_{1-x}$Ti$_x$)S$_3$ for x = 0, 0.01, 0.02, 0.03, and 0.04 derived from absorption spectrum. As can be seen from Figure 4(b), the extracted band gap for Ba(Zr$_{1-x}$Ti$_x$)S$_3$ (x = 0) is 1.78 eV, in good agreement with our previous results.[15, 26] With increasing Ti concentration, the band gap decreases monotonically. With a mere 4 at% Ti alloying, the band gap has dropped to 1.51 eV. It has been shown in earlier theoretical work that the conduction band minimum and valence band maximum of chalcogenide perovskites mainly consist of transition metal d-states and chalcogen p states, and there is little intermixing of these states.[27] By substitutional alloying of Ti, the 4d states of Zr are partially replaced by the 3d states of Ti, which are significantly deeper in energy. This then down shifts the conduction band minimum and narrows the band gap. Experimental results from $Ba(Zr_{1-x}Ti_x)S_3$ with higher Ti concentration are supplied in supporting information Figure S5. However, it can be seen that in addition to the redshift in energy, there is also a significant absorption tail at the low energy end. This absorption tail may be attributed to the secondary phases. For example, $BaTiS_3$ is known to crystallize in the hexagonal phase with a small band gap of 0.2 eV.[27, 31]

The XRD and Raman spectra confirm pure perovskite phase in Ti-alloyed $Ba(Zr_{1-x}Ti_x)S_3$ systems at low alloying concentration of x ≤ 0.04. The UV-vis absorption suggests that despite the low alloying concentration of just 4 at%, the band gap is effectively reduced by 270 meV, making it close to the optimal band gap to approach the Shockley-Queisser limit. Since the thermodynamically stable structures of $BaZrS_3$ and $BaTiS_3$ are very different with different band gaps and optical absorption properties, it is important to understand the structural stability of the material by alloying Ti into $BaZrS_3$. To explore the stability, we further measured and compared the XRD patterns for 0 at% and 4 at% Ti alloyed samples. As can be seen from Figures 5(a) and (b), the XRD patterns of both 0 at% and 4 at% Ti alloyed samples show no measurable difference for as-synthesized samples and those prepared 890 days ago and stored at ambient atmosphere, suggesting that Ti alloying does not affect its long term stability against room temperature oxidation or moisture. Previous research has verified the stability of $BaZrS_3$ under the temperature of <550 °C in air.[32] In the present work, the 0 at% and 4 at% Ti alloyed $Ba(Zr_{1-x}Ti_x)S_3$ powder samples were annealed under partial $O_2$ atmosphere for 1 hour at 500 °C and 800 °C, respectively.

As can be seen from Figures 5(a) and 5(b), both samples show no obvious secondary phases after one hour annealing at 500 °C, showing the stability of Ba(Zr$_{1-x}$Ti$_x$)S$_3$ system against oxidation at below 500 °C. Upon one hour annealing at 800 °C, the major phase of the unalloyed sample remains to be the distorted perovskite phase, with the formation of small amount of secondary phases, estimated to be < 10% from the integrated peak intensity. However, under the same condition, the 4 at% Ti alloyed sample has decomposed into BaS and ZrO$_2$ (as shown by the asterisk marks), with only a small percentage of BaZrS$_3$ phase. These results suggest that Ti-alloyed BaZrS$_3$ is stable up to 500 °C, and higher temperature leads to its decomposition. As a result, Ti incorporation into BaZrS$_3$ does lead to a distorted perovskite Ba(Zr$_{1-x}$Ti$_x$)S$_3$ structure that is less stable than its parent compound.

## CONCLUSION

In conclusion, we have synthesized Ti alloyed Ba(Zr$_{1-x}$Ti$_x$)S$_3$ powders with x from 0 to 0.1. The materials show good crystallinity, stability, and optical properties. Both the XRD and Raman spectroscopies suggest successful alloying of BaZrS$_3$ by Ti via substitutional alloying. At lower alloying concentration ≤ 4 at%, the band gap of Ba(Zr$_{1-x}$Ti$_x$)S$_3$ is effectively reduced. With merely 4 at% Ti alloying, the band gap of Ba(Zr$_{1-x}$Ti$_x$)S$_3$ can be tuned from 1.78 eV to 1.51 eV, allowing a theoretical PCE of 32% for solar cells by using Ba(Zr$_{1-x}$Ti$_x$)S$_3$ as the light absorber. Higher Ti-alloying concentration (>5 at%), on the other hand, is found to destabilize the distorted chalcogenide perovskite phase. Importantly, our work on band gap engineering by Ti alloying of BaZrS$_3$ will also merit future studies of thin films for solar cell devices.

## EXPERIMENTAL

BaCO$_3$, ZrO$_2$ and TiO$_2$ powder were mixed together and ball milled at 40 rpm for 20 hours using an MTI MSK-SFM-1 ball mill with four stainless steel milling jars. The mass of BaCO$_3$ powder was 3.947g, while the mass of ZrO$_2$ and TiO$_2$ varied from 2.464 g to 2.218 g, and 0 to 0.160 g, respectively, to obtain different Ti alloying level from 0 at% to 10 at%. All precursors were from Alfa Aesar with a purity of >99.8%. The mixture was then heated up to 1000 °C with a heating and cooling rates of 3°C/min, and kept at 1000 °C for 8 hours in air using an MTI GSL-1700X tube furnace. The above procedures were repeated to obtain Ba(Zr$_{1-x}$Ti$_x$)O$_3$ powder samples. Sulfurization procedure was performed in an MTI OTF-1200X tube furnace using CS$_2$ as the sulfur source at a temperature of 1050 °C for 4 hours. The CS$_2$ flow was optimized to be ~20 standard cubic centimeters per minute (sccm) and was turned on as the temperature was ramped up to 800 °C and turned off as the temperature was ramped down to 800 °C. 1M NaOH solution was used to remove unreacted $CS_2$ during sulfurization. Sulfurization details can be found in our previous published work.[15]

XRD results were obtained through a Rigaku Ultima IV X-ray diffraction system with an operational X-ray tube power of 1.76 kW (40 kV, 44 mA). The powder XRD measurements were performed under θ-2θ scanning mode and continuous scanning type with a step size of 0.01°. Room temperature Raman spectra were measured using a Renishaw inVia Raman Microscope with a 1200 l/mm grating, 50× objective lens, and 514 nm laser. A Labsphere RSA-HP-8453 diffuse reflectance spectroscopy accessory attached to Agilent 8453 ultra-violet/visible (UV-vis) spectroscopy system was used to obtain the absorption spectrum. Surface morphology and energy-dispersive X-ray elemental analysis were performed using a Focused Ion Beam-Scanning Electron Microscope (FIB-SEM) – Carl Zeiss AURIGA CrossBeam with an Oxford EDS system.

## ASSOCIATED CONTENT

## Supporting Information

XRD patterns of $Ba(Zr_{1-x}Ti_x)O_3$ powders; SEM images of 0 to 10 at% Ti alloyed $Ba(Zr_{1-x}Ti_x)S_3$ powders; The XRD patterns, Raman spectra, and Tauc Plot of 5 to 10 at% Ti alloyed $Ba(Zr_{1-x}Ti_x)S_3$ powders;

## AUTHOR INFORMATION

### Corresponding Author

*E-mail: haozeng@buffalo.edu.

### Notes

The authors declare no competing financial interest.

**ACKNOWLEDGEMENT**

Work supported by US NSF CBET-1510121, CBET-1510948, MRI-1229208, and DOE DE-EE0007364. Y. -Y.S. acknowledges support by NSFC under Grant 11774365. We thank Bernard Weinstein for fruitful discussions.

## Figure captions

**Figure 1.** Typical SEM images of Ba(Zr$_{1-x}$Ti$_x$)S$_3$ powder samples for (a) x = 0 and (b) x = 0.04; and the EDX spectra of Ba(Zr$_{1-x}$Ti$_x$)S$_3$ for (c) x = 0 and (d) 0.04. The insets are enlarged view of the characteristic emission peak at around 4.5 keV.

**Figures 2.** (a) XRD patterns of Ba(Zr$_{1-x}$Ti$_x$)S$_3$ powder samples for x = 0, 0.01, 0.02, 0.03, and 0.04. The enlarged view of (b) (121) peak; (c) (040) peak, and (d) (240) peak of Ba(Zr$_{1-x}$Ti$_x$)S$_3$. All peaks shift to higher angles with increase of x from 0 to 0.04. (e) The d spacing as a function of Ti alloying percentage calculated from the position of the (121) peak. The decrease in d spacing is consistent with substitutional alloying of Ti with atomic radius smaller than that of Zr.

**Figure 3.** The Raman spectra of Ba(Zr$_{1-x}$Ti$_x$)S$_3$ for x = 0, 0.01, 0.02, 0.03, and 0.04, measured at room temperature.

**Figure 4.** (a) The Tauc plot of Ba(Zr$_{1-x}$Ti$_x$)S$_3$ powder samples for x = 0, 0.01, 0.02, 0.03, and 0.04. A linear fitting shown by the dashed lines indicates the band gap energy of 1.78 eV and 1.51 eV for 0 and 4 at% alloyed samples, respectively; (b) Band gap energy as a function of Ti alloying percentage.

**Figure 5.** XRD patterns of (a) 0 at% and (b) 4 at% Ti alloyed Ba(Zr$_{1-x}$Ti$_x$)S$_3$ powder samples annealed at 500 and 800 °C, together with those of as-synthesized samples and samples stored at ambient conditions for 890 days. Black and red asterisk marks in (a) and (b) indicate standard XRD peak positions of BaS and ZrO$_2$, respectively. All other unlabeled peaks belong to BaZrS$_3$.

**Figures**

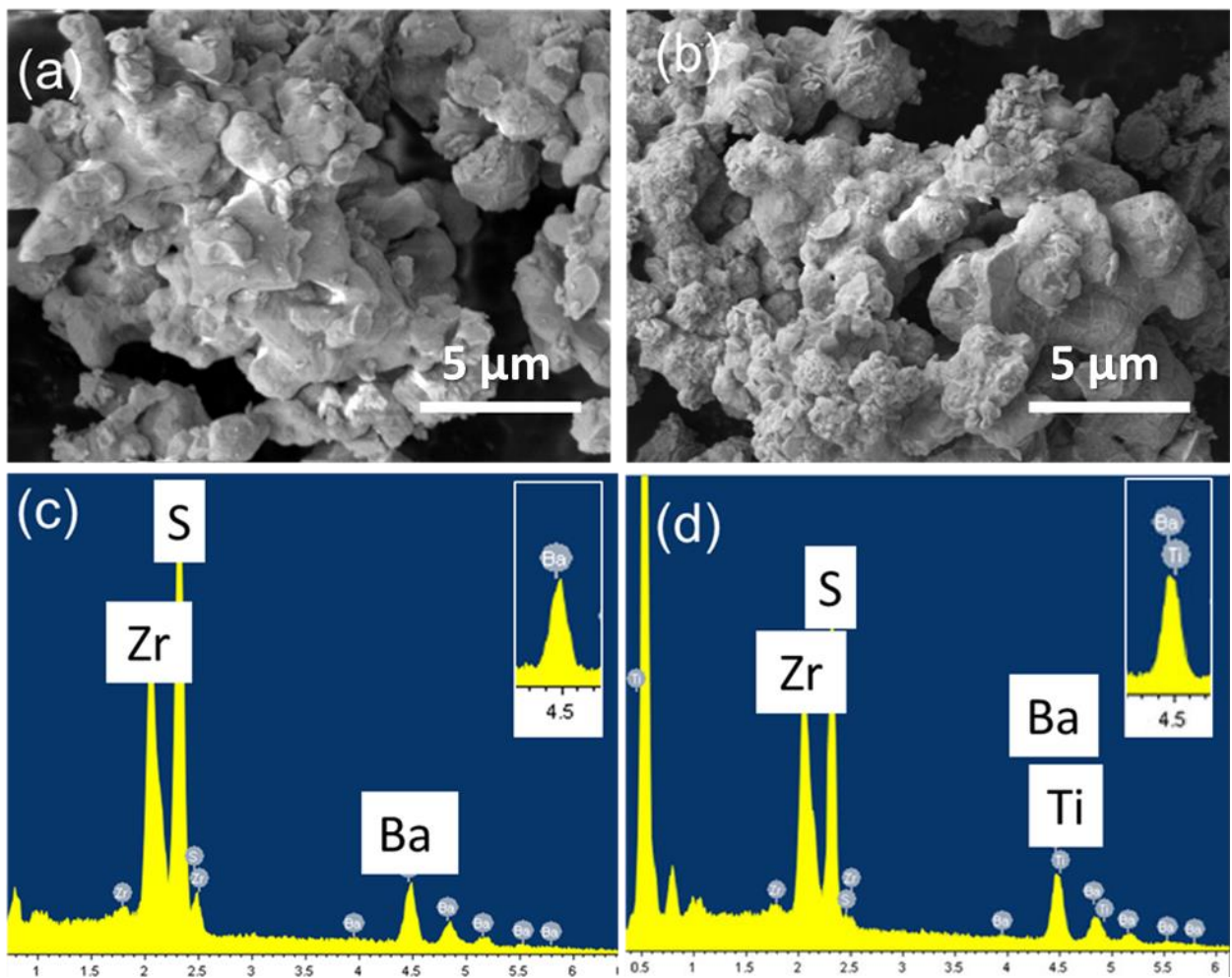

**Figure 1**

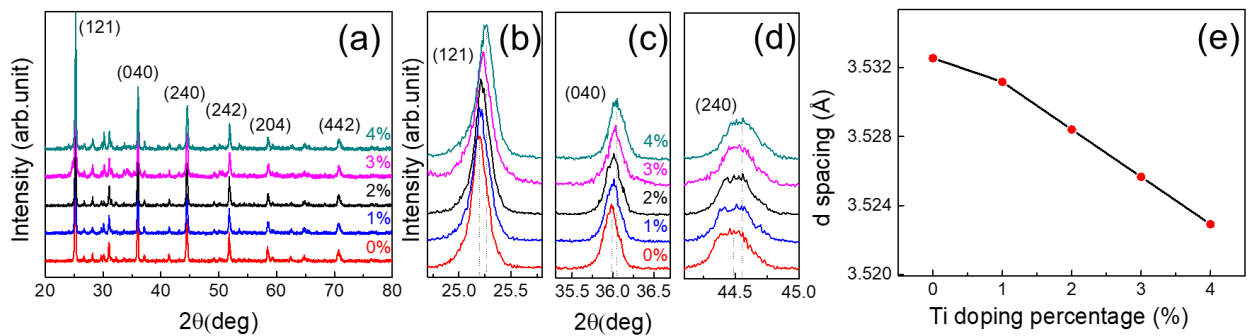

**Figure 2**

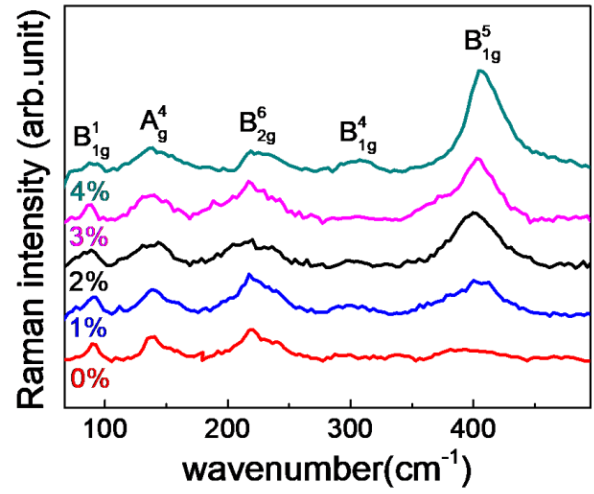

**Figure 3**

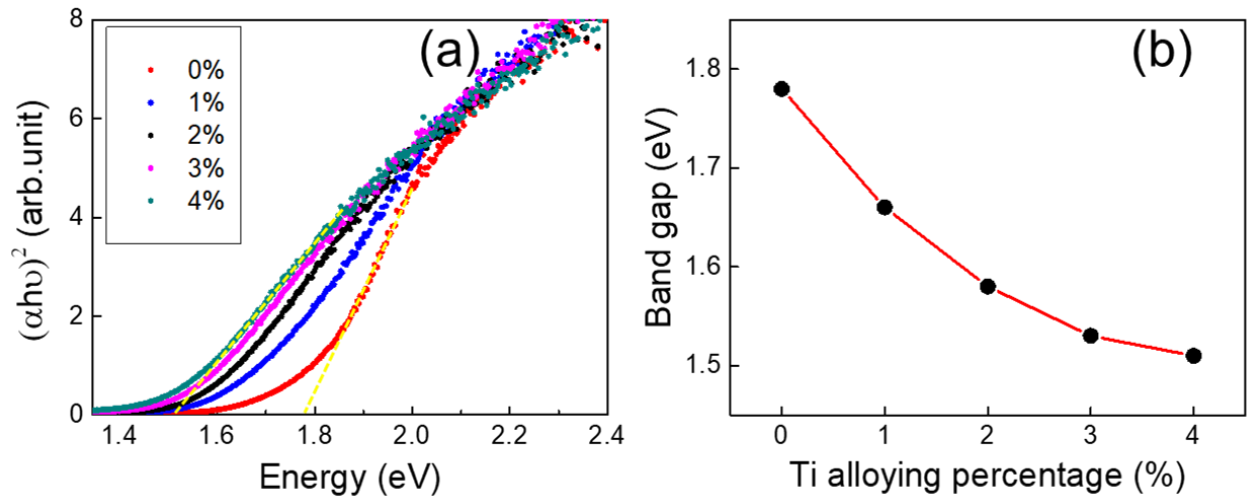

**Figure 4**

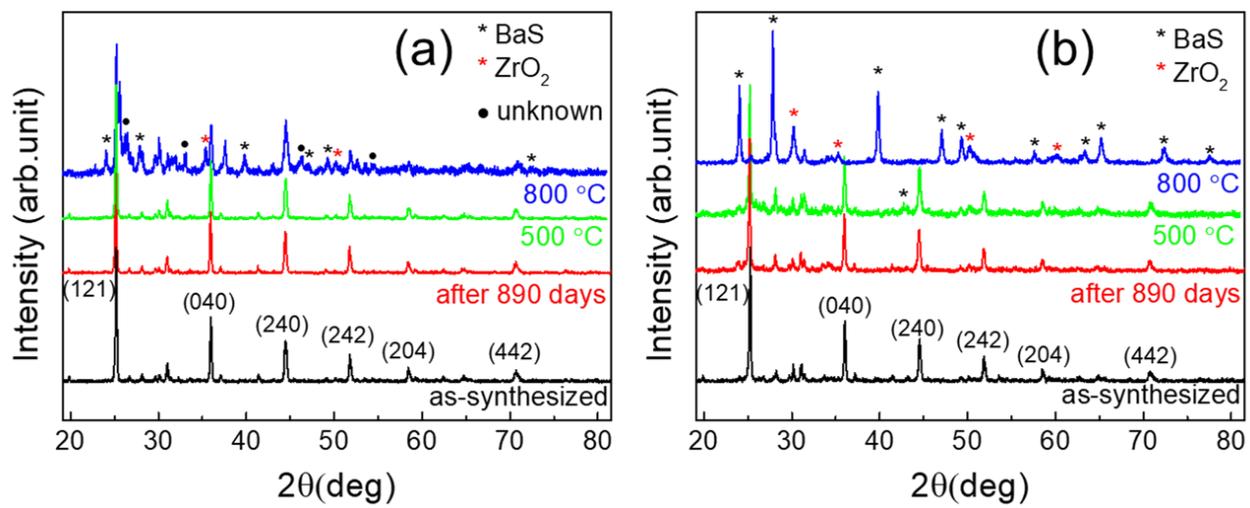

**Figure 5**